\documentclass{sig-alternate}
\usepackage{cite}
\usepackage{graphicx}
\usepackage{subfigure}
\RequirePackage{epsfig}
\usepackage{xspace}
\usepackage[linesnumbered,ruled,vlined]{algorithm2e}
\usepackage{algorithmic}

\usepackage{amsmath}
\usepackage{amssymb}
\usepackage{amscd}
\usepackage{setspace}
\usepackage{enumerate}

 \newtheorem{theorem}{Theorem}[section]

    \newtheorem{definition}[theorem]{Definition}

\begin{document}

\conferenceinfo{ISSAC}{'13 Boston, USA}
\CopyrightYear{2013} 

\title{\LARGE \bf A Preprocessor Based on Clause Normal Forms and Virtual Substitutions to Parallelize \\Cylindrical Algebraic Decomposition}

\numberofauthors{1} 
\author{
\alignauthor
Hari Krishna Malladi ~~ and ~~ Ambedkar Dukkipati\\
       \affaddr{Indian Institute of Science,}
       \affaddr{India}\\
 \email{\{harikrishnamalladi,ad\}@csa.iisc.ernet.in}
}


\date{31 December 2011}
\maketitle
\thispagestyle{empty}

\begin{abstract}
The Cylindrical Algebraic Decomposition (CAD) algorithm is a comprehensive tool to perform quantifier elimination over real closed fields. CAD has doubly exponential running time, making it infeasible for practical purposes. We propose to use the notions of clause normal forms and virtual substitutions to develop a preprocessor for CAD, that will enable an input-level parallelism. We study the performance of CAD in the presence of the preprocessor by extensive experimentation. Since parallelizability of CAD depends on the structure of given prenex formula, we introduce some structural notions to study the performance of CAD with the proposed preprocessor.\end{abstract}

\terms{Algorithms, Performance, Theory}

\date{8 June 2012}
\maketitle
\thispagestyle{empty}
\section{Introduction}

The study of real algebraic geometry deals with the study of the real roots of an equation as, more often than not, the real roots are the most desired solutions to a system of equations. It is important to note that for some problems there are no counterparts in complex algebraic geometry.
Given a first-order formula in real algebraic geometry with both quantified and quantifier-free variables, the process of finding an equivalent first-order formula in these quantifier-free variables is called quantifier elimination. When the first-order formula is a boolean combination of polynomial equations and inequalities, we can consider quantifier elimination as a problem in real algebraic geometry. 

The algorithm specified by Tarski~\cite{tarski} for quantifier elimination is highly resource intensive. So newer and more efficient algorithms have come up and replaced it. The fact that projections upon parameters of a semialgebraic set (semialgebraic set is defined as the solution space of polynomial equations and inequalities) are also semialgebraic (Tarski-Seidenberg principle) has led to Cylindrical Algebraic Decomposition (CAD)~\cite{collins}, which eventually became the standard algorithm for quantifier elimination. The time complexity of CAD algorithm is doubly exponential in the number of variables (both quantified and quantifier-free). 

CAD is a recursive algorithm, which constitutes of a sequence of projections, followed by a sequence of constructions. It has undergone an extensive array of developments, such as Hong's Partial CAD~\cite{partialcad}, Hong's projection operator~\cite{hongproj}, Scott McCallum's projection operator~\cite{mccallumproj}, etc. An application of Gr\"{o}bner Bases to CAD to improve the time complexity is studied in~\cite{qegrob}. There were also some efforts to parallelize CAD~\cite{issac89}. In spite of all these improvements, the fact that CAD's time complexity is doubly exponential makes quantifier elimination through CAD impossible for a wide range of real-world applications.

Parallelism has appeared to be the trend, while addressing algorithms which have a significant amount of independence in their structure. Speeding up executions by scaling the existing CPU speeds has been found to be inefficient. The maximum possible clock speed has remained at around 3 GHz for the past 12 years, for a desktop microprocessor, in spite of being scaled by about 100 times in the 1990s. Multicores have achieved prominence, as dual-cores and quad-cores have become ubiquitous in the past 7 years. This suggests that the only major way to improve an existing algorithm's running time is through parallelism. We use this as a motivation to introduce an input-level parallelism in the CAD algorithm.

In this paper, we study a possibility to preprocess the input prenex formula, so that several instances of CAD can be run in parallel on it. Input-level parallelism is not directly applicable to CAD because of the constraints imposed by first-order logic itself. Thus, clause normal forms and virtual substitutions are used to `separate' a given input formula so that it's components can be executed in parallel using existing CAD implementations. We study the preprocessor algorithm through extensive experimentation using the QEPCAD B tool, running several instances on a cluster.

The work by Saunders et. al.~\cite{issac89} has brought up the notion of a parallel version of CAD, where they introduce an execution-level parallelism to make the phases of CAD work in parallel. Since our work introduced input-level parallelism, our preprocessor can be used along with Saunders' approach.

The paper is organized as follows. The preliminary notations and the original CAD algorithm are briefly introduced in Section 2. The use of clause normal forms and virtual substitutions and the preprocessor itself are described in Section 3. In Section 4, we introduce some notions to define structure in a prenex formula. An experimental characterization of the preprocessor algorithm, are presented in Section 5. Extensions sought for the proposed algorithm and a few concluding remarks have been made in Section 6.

\section{Preliminaries}

\subsection{Basic Definitions and Notation}

Let $k [x_1,. . .,x_n]$ denote the set of all polynomials in variables, $ x_1,. . .,x_n $ and coefficients from the field $k $. The base field, $k $ is assumed to be the set of real numbers, $\mathbb{R}$ throughout the paper.

\begin{definition} A formula of the form \begin{displaymath}(Q_{1}x_{1}),. . .,(Q_{n}x_{n}) [\psi(y_{1},. . .,y_{m},x_{1},. . .,x_{n})]\end{displaymath} is called a \textit{prenex} formula, where $Q_{i} \in \{ \exists, \forall \}$ and $\psi$ is a quantifier-free formula in $x_1,,. . .,x_n,y_1,. . .,y_m$.
\end{definition}

\begin{definition} $V \subset \mathbb{R}$ is said to be a semialgebraic set if there exists $ f_1,. . .,f_s \in \mathbb{R} [x_1,. . .,x_n]$ and $g_1,. . .,g_t \in \mathbb{R} [x_1,. . .,x_n]$ such that $V = \{ (a_1,. . .,a_n) \in \mathbb{R} : f_i(a_1,. . .,a_n) = 0, i=1,. . . ,s, g_j(a_1,. . .,a_n) \geq 0, i=1,. . .,t\}$.
\end{definition}

 A connected subset of $\mathbb{R}^{n}$ is called a `region' in $\mathbb{R}^{n}$. Consider a region $R$ and functions $f_i : \mathbb{R}^n \longrightarrow \mathbb{R}, i=1,. . .,l$ satisfying, $f_{1} < f_{2} < ... < f_{l}$. This ordering ensures that these functions do not intersect each other. Graph of $f_{i}$ is called `$f_{i}$-section'. In other words, `$f_{i}$-section is the set containing all points of the form $(a,b)$, where a $\in \mathbb{R}^{n-1}$ and $b = f(a)$. An `(f$_{i}$,f$_{i+1}$)-sector' is the set of all points $(a,b)$, where a $\in \mathbb{R}^{n-1}$ and $f_{i}(a)$ $< b <$ $f_{i+1}(a)$. A `cylinder' over a region $R$ is the set of all points $(a,b)$, where $a \in R$ and $b \in \mathbb{R}$. A `stack' over a region $R$ is the collection of sections and sectors that occur in the cylinder over $R$. A partition of $\mathbb{R}^{n}$ into semialgebraic partitions is called an `algebraic decomposition' of $\mathbb{R}^{n}$.

\begin{definition} An algebraic partition of $\mathbb{R}^{n}$ satisfying the following two properties is called a `cylindrical algebraic decomposition':
\begin{enumerate}
 \item If $n=1$, then the CAD is a set of points and open intervals.
 \item If $n>1$, every region of the CAD of $\mathbb{R}^{n-1}$ has a stack over it, which is a disjoint subset of the CAD of $\mathbb{R}^{n}$.
\end{enumerate}
\end{definition}

 A set of polynomials, $F$ is said to be `sign invariant' on the region $R$ iff no polynomial in $F$ changes its sign anywhere on $R$.

\begin{definition} A CAD $C$ of $\mathbb{R}^n$ is sign invariant with respect to a set of polynomials, $P$ if and only if $P$ is sign invariant on every cell of $C$.
\end{definition}

The algorithm to generate the CAD of $\mathbb{R}^{n}$(which shall be referred to as `CAD Algorithm') has two phases of execution, namely
projection and construction. Projection is specified by a projection operator, whereas construction depends on the projection phase.

\subsection{Cylindrical Algebraic Decomposition Algorithm}

CAD takes as input, a set of polynomials in $n$ variables and generates a cylindrical algebraic decomposition
of the $n$-dimensional real space. This algorithm works recursively to produce a series of projections on lesser dimensions and builds the CAD of each dimension while returning the recurring functions. An outline of CAD algorithm~\cite{20yrs} is presented below. The intricate details have been omitted as the purpose of this presentation is to provide a glimpse into the algorithm, rather than to study the algorithm itself.

\begin{enumerate}
\item Project each polynomial onto lesser dimensions. Polynomials in $n$ variables are taken as input and a set of polynomials in $n-1$ variables is obtained as output, upon applying a projection operator. This process is continued till a set of univariate polynomials is obtained.
\item These univariate polynomials can easily be solved (using Sturm's theorem iteratively or otherwise). The roots of these univariate polynomials and the intervals between them are taken as the regions in the CAD of one dimensional real space. Designate a point in each region as a sample point.
\item For the two dimensional CAD, substitute the sample point found in each one-dimensional region, $R$ in each of the polynomials in the 2-dimensional projection to again get univariate polynomials. Find the roots of these polynomials and create two dimensional regions that form the stack over the region $R$. Designate sample points for these regions as well. Suppose we are given a CAD for $k$-dimensional real space, we use the projected set for $k+1$-dimensions and substitute the sample points of the $k$-dimensional CAD in them to get univariate polynomials. These can now be solved to get $k+1$-dimensional CAD.
\end{enumerate}

Now, this information can be used by Algorithm 1 to perform quantifier elimination~\cite{andreas}. Assume that the variables $x_1,...,x_k$ are quantifier-free and $x_{k+1},...,x_n$ are quantified, and let $Q_i$ be the corresponding quantifier for $i\in\{k+1,...,n\}$.

\begin{algorithm}
 \caption{Algorithm for Quantifier Elimination}

 \KwIn{The CADs of $\mathbb{R}^{1}$ to $\mathbb{R}^{n}$ $(D_{1},. . .,D_{n})$, the formulae describing the regions of the CAD of $\mathbb{R}^{k}$ $(F_{1},. . . ,F_{l}$) and  the input formula (whose quantifiers are to be eliminated) ($\psi $).}
 \KwOut{The quantifier-free formula equivalent to the input prenex formula.}
\begin{enumerate}
 \item For $k \leq i < n$, we have
  \begin{enumerate}
  \item If $Q_{i+1}$ is $\exists$, then a cell,$C \in D_{i}$ is valid if at least one cell in the stack over $C$ is valid.
  \item If $Q_{i+1}$ is $\forall$, then a cell,$C \in D_{i}$ is valid if all cells in the stack over $C$ are valid.
  \end{enumerate}
 \item A region, $C$ in $D_{n}$ is valid if $\psi(t_{C})$ is TRUE, where $t_{C}$ is the sample point of that cell.
 \item Obtain the cells of $D_{k}$ which are TRUE. The disjunction of the formulae of these cells is the required quantifier-free formula and is returned.
\end{enumerate}
\end{algorithm}

\section{Towards Parallelization}

\subsection{Using Clause Normal Forms}

In this section we introduce the application of miniscoping and clause normal forms to obtain a preprocessor that can parallelize CAD. Miniscoping is prevalent in the standard literature as a way of localising quantifiers. One can use an algorithm by Nonnengart and Weidenbach~\cite{miniscoping} to compute `clause normal forms'. 

\begin{definition} A sentence $\phi$ is said to be in `Clause Normal Form', if $\phi = \forall x_1,. . . ,\forall x_{k}[C_1 \wedge . . . \wedge C_{k}]$ where $C_{i} = L_{i,1} \vee . . . \vee L_{i,l_{i}}$ and each $L_{i,j}$ is a literal. Clause normal forms shall be represented by `CNF' throughout this paper.\end{definition}

Their work introduces this algorithm and proves that it terminates in a finite amount of time and that the generated clause normal form is equivalent to the input. The time complexity of this algorithm is polynomial in the number of statements.

The algorithm to compute clause normal forms uses the concept of skolemization to eliminate existential quantifiers (by systematic replacement of existentially quantified variables by functions of universally quantified variables). Skolemization may introduce skolem functors of high exponents, but the doubly exponential running time of CAD enables us to cope with them in practice, due to a reduction in the number of variables (brought about by skolemization).

CNFs are used to impose a structure on the input prenex formulae, that will make computation easier. But, a well-crafted transformation on a prenex formula may result in a form which performs better than the CNF case. But, such well-crafted transformations usually require a great amount of computational power (and possibly non-determinism) to achieve in practical scenarios. Thus, the CNF has been used to give a measure of speed-up that can be achieved in an average case, as the CNF can be obtained using a deterministic algorithm, which terminates for any input.

\subsection{Using Virtual Substitutions}

The CAD algorithm benefits from substitutions, which might reduce the number of variables in some clauses, prior to the separation phase. This may decrease the dependence of one clause on another. But substitutions can be non-trivial and in many cases, impossible. Substitutions can be used with linear and quadratic equations and inequalities directly. For cubic and quartic polynomials, substitutions can be used only if it is obtained easily. Newton and Cardano's formulas for quartics and cubics involve complex roots, and hence cannot be used. For quintic and higher order polynomials, there exists no generic closed form solution, as proved by Abel.

We would like to use the algorithm to perform virtual substitutions by Volker Weispfenning~\cite{weispfenning}. While Weispfenning's work on virtual substitution seeks to find a quantifier-free equivalent by itself, we would like to adapt it to minimize dependencies within the clauses and hence, to increase parallelism.

In our notion of substitutability, we classify polynomials into two classes, namely (i) substitutor and (ii) substituend.

A `substitutor' polynomial is used to substitute for a variable in other polynomials. A `substituend' polynomial is one, in which a variable is replaced.Assume that $P$ is any polynomial and $T$ is a substitutor. We say that $T$ is `substitutable' in $P$ if, the process of replacing a variable in $P$ by virtual substitution results in a decrease in the number of variables in the polynomial $P$.

We use the results proved by Weispfenning and Christopher Brown~\cite{brownsub}, which state that virtual substitution leads to an equi-satisfiale formula.

\subsection{The Proposed Preprocessor}

In this section we present a preprocessor to the CAD algorithm that is motivated by the notion of clause normal forms and virtual substitutions.

Virtual substitutions would be performed prior to clause normal form computation, as substitutions might result in an increase in the number of clauses.

The clause normal form has been chosen as the preferred format for computational efficiency. The fact that any given prenex formula can be converted to a clause normal form enables us to perform the experimental evaluation on test cases containing solely of clause normal forms.

We now list the proposed algorithm.

\begin{algorithm}
 \KwIn{A prenex formula containing a boolean combination of polynomial equations and inequalities, $f$}
 \KwOut{Quantifier-free equivalent of $f$. }
\begin{enumerate}
\item Compute the negation normal form of $f$ and call it $f'$.
\item Perform virtual substitutions after identifying substitutors and substituends and call the resulting formula, $f''$.
\item Compute the clause normal form of $f''$ and call it $f'''$.
\item Call an instance of CAD for each clause in $f'''$, with each instance being an independent thread of execution.
\item Concatenate the outputs of the $k$ instances of CAD using boolean conjunction.
\end{enumerate}
 \caption{The Preprocessor Algorithm}
\end{algorithm}



\section{Structural Notions of the Prenex Formulae and the Preprocessor}

The introduction of CNF computation introduces the notion of partitioning the space of all prenex formulae into classes, depending on how many clauses it could be separated into. To interpret the results presented above, we use the following definitions.

A prenex formula is said to be separable if it can be split into two or more clauses. Separability is a property of the prenex formula. Separability may result in the loss of structure in a given prenex formula. By this we mean that the interdependencies among the formula's constituent polynomials might be lost, resulting in wasteful and redundant computations, which would not have been required if the formula was not separated. We need a formal, quantitative definition of `structure' with respect to a given formula to study it's effects on the running time of the proposed algorithm.

\begin{definition}
The `sharing factor' existing among two given prenex formulae, $f$ and $g$, is the number of variables that are shared between $f$ and $g$. It's denoted by $T_{f,g}$.
\end{definition}

The sharing factor essentially captures the distribution of variables among the constituent polynomials of a prenex formula. If formulae $f$ and $g$ have $k$ variables in common, it implies that $f$ imposes conditions on $g$ in $k$ dimensions. The sharing factor gives a measure of structure present in a formula, as the sharing of variables between polynomials causes interdependencies among them. Having defined the sharing factor, we now define the operation, decomposition, where the sharing of variables comes into consideration.

\begin{definition}
Consider a prenex formula, $f$, which has $n$ variables, out of which, there are $k$ quantified variables and $n-k$ quantifier-free variables. We say that the decomposition of $f$, $F$ is the set consisting of all the clauses which occur in the CNF of $f$.
\end{definition}

This definition of decomposition defines separability as a property of the formula. In the remainder of this section, we assume that all prenex formulae are in variables, $x_1,. . .,x_n$. It is natural to classify prenex formulae into classes based on the number of clauses it can be separated into. Thus, we have the following definition.

\begin{definition}
A prenex formula is defined to be k-separable if its CNF has at least $k$ clauses. The set of all k-separable formulae shall be denoted by $S_k$.
\end{definition}

According to this definition, we can observe that $S_k \subset S_{k-1}$. The set of 1-separable formulae is the set of all prenex formulae, which shall be denoted by $P$. We assume that we have only a constant number of processors, and we denote this by $K$. This leads to a definition of the class of separable problems.

\begin{definition}
Consider a set $S \subset P$. $S$ is called the `separable class' if the following conditions hold.
\begin{itemize}
\item $S \subset S_2$
\item If $f$ is a prenex formula and $F$ is it's decomposition, $\forall f \in F$, $\exists x_i \in \{x_1,...,x_n\}$ such that $f$ is independent of $x_i$.
\end{itemize}
\end{definition}

The property of separability is not uniform in this separable class. It depends heavily on the sharing factor of the formula. The quantification of the term structure enables us to define the `centre of $S_k$' for $k < K$.

\begin{definition}
The centre of $S_k$ is the set of all prenex formulae, $f$ in $n$ variables ($x_1, . . . ,x_n$) with decomposition $F$, such that the clause $f_i \in F$ contains exactly the variables $x_{\frac{i-1}{k}+1}, . . . ,x_{\frac{i}{k}}$.
\end{definition}

Unlike the classes $S_1, . . . ,S_k$, the classes $C_1, . . . ,C_k$ do not form a nesting chain structure. In other words, $C_i$ need not contain $C_{i+1}$.

This centre of a separable class, $S_i$, is the set of formulae which pose a `best case' scenario for the CAD algorithm to run in parallel for an input prenex formula containing $i$ variables. The lower the maximum sharing factor in a prenex formula, the faster it's execution would be on the cluster, using the parallel CAD algorithm. This has been demonstrated by considering prenex formulae with zero sharing factor.

But, as the sharing factor between two clauses increases, we find that these two clauses become interdependent to a greater extent. Thus, executing them in parallel would lead to greater amount of work being done, as compared to the case where the conditions imposed by one clause influences the computation of CAD on another clause.

\section{Experimental Evaluation}

The simulation results in this paper utilize the clause normal form (CNF), to ease the process of generating prenex formulae. The test cases are always in the CNF. The simulations are performed on a cluster, comprising of 36 nodes, with 8 Intel Xeon quad-core CPUs in each node. The preprocessor is implemented using C and the OpenMP library. The code used for these experiments has been uploaded at http://algoalgebra.csa.iisc.ernet.in/Preprocessor/. 

The experimental setup is as follows:
\begin{enumerate}[(i)]
\item The dependence of running time of QEPCAD B on number of terms, highest exponent and total number of terms in a prenex formula,
\item The distribution of prenex formulae according to the number of clauses in their corresponding CNFs,
\item The comparison of running times of QEPCAD B, with and without the preprocessor and
\item The structure imposed be dependence among polynomials is studied in terms of the sharing factor.
\end{enumerate}
\subsection{On the Parameters of CAD}

Maximum number of terms per polynomial and the highest exponent across all polynomials in a CNF are two of the major parameters which influence the running time of the CAD algorithm. As there is no closed expression for running time of CAD in terms of these parameters, it seems appropriate to plot the variations of these parameters versus the running time. The values presented are averaged over 100 randomly generated prenex formulae. Each prenex formula that is generated, is a CNF, where each clause consists of a set of polynomial equations or inequalities separated by the boolean disjunction. The default random number generator, \texttt{rand()} was used.\\\\

\begin{figure}
\begin{center}
  \includegraphics[scale=0.3]{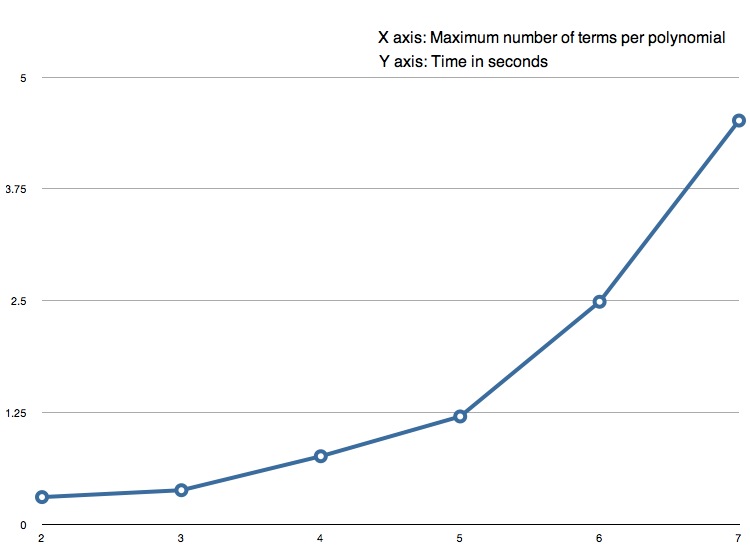}\\
\caption{Running Time vs Max. No. of Terms plot.}

  \includegraphics[scale=0.3]{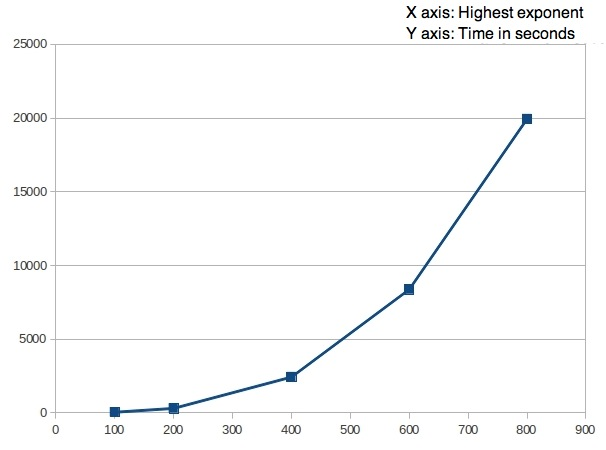}\\
\caption{Running Time vs. Highest Exponent plot for 2 variables.}

  \includegraphics[scale=0.3]{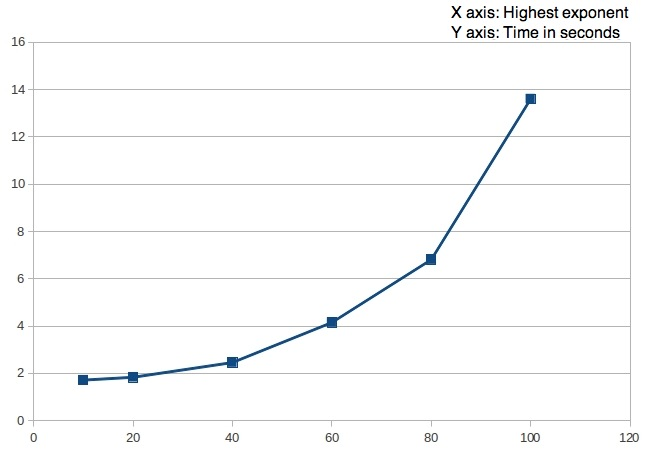}\\
\caption{Running Time vs. Highest Exponent plot for 3 variables.}

  \includegraphics[scale=0.3]{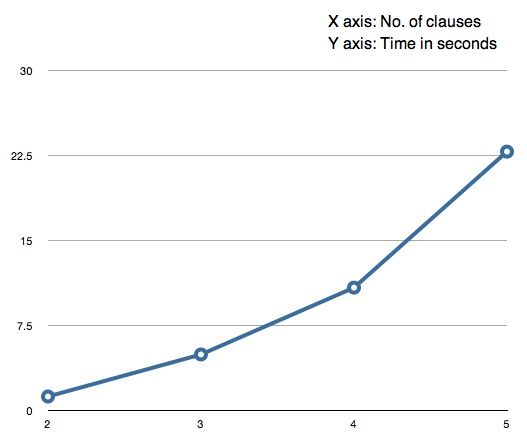}\\
\caption{Running Time vs. Total number of terms.}
\end{center}
\end{figure}

First we fix the maximum number of terms, say T. The number of terms in each polynomial is a random number in between $0$ and $T$. The number of clauses, the highest exponent and number of variables in the prenex formula were kept constant. The increase in running time was observed to be close to exponential in this case, as depicted in Fig. 1.\\

This increase is not surprising as an increase in the number of terms per polynomial results in an exponential increase in the total number of terms that are processed, throughout the course of execution. This is due to an increase in number of terms in each set of projected polynomials.


The highest exponent, $E$, was kept constant in each stage of the experiment and all other exponents were taken as random numbers between $0$ and $E$. All other parameters such as number of clauses, maximum number of terms and the number of variables were kept constant. The experiment was performed for both, a 2 variable case and a 3 variable case. The increase in the running time was observed to closely mimic Fig. 1, as shown in Fig. 2 and Fig. 3. \\

This increase is expected as an increase in the exponent results in an increase in the number of regions, when the construction phase of CAD reaches the variable with the exponent $E$. This is leads to an almost exponential increase in the number of regions, as each previously generated region spawns more regions in the higher dimensions. This would result in a stark increase in running time.

One can conclude that number of terms and exponents contribute equally to the running time of CAD.


This was conducted for a 2 variable case, with a maximum of $5$ terms per polynomial and a maximum of 2 polynomials per clause. The number of clauses was varied from $2$ to $6$. The results were averaged over 50 randomly generated prenex formulae. An almost exponential increase is observed in this case, as shown in Fig. 4. 

One could infer from this decrease in gradient that the total number of terms in a collection of projected polynomials depends to a greater extent on the number of terms per polynomial, than on the number of polynomials.

\subsection{On Validity of the Preprocessor}

As the analysis is performed on the assumption that the input is in the CNF format, with each prenex formula containing two or more clauses, it is necessary to give an account on the percentage of randomly generated prenex formulae, that can be converted to a CNF containing more than one clause. A CNF is a conjunctive normal form with universally quantified variables. Hence, 100 prenex formulae were randomly generated, which do not conform to the normal form defined previously. They were subjected to a logic converter, which converts them to a minimal conjunctive normal form. We have observed the following from the plot in Fig. 5: only 8\% of the formulae were 1-separable. Hence, in 92\% of the cases, the algorithm would lead to a parallel execution of CAD.\\
\begin{figure*}
\begin{center}
  \includegraphics[scale=0.5]{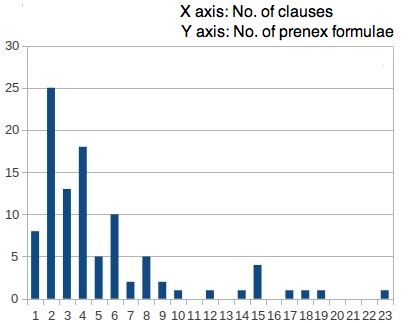}\\
\caption{Distribution of the number of clauses in the CNF version of a randomly generated formula.}
\end{center}
\end{figure*}

\subsection{On the Performance of CAD with the Proposed Preprocessor}

With the above two analyses in place, the next step is to experimentally compare the running times of an implementation of CAD algorithm with and without the preprocessor. The analysis is done for 2,3 and 4 variable cases. In each case, the prenex formula is 100-separable and is in Clause Normal Form. Each clause has at most 5 polynomials and each polynomial has at most 5 terms. The analysis is done for the 2, 3 and the 4 variable case. The implementation uses OpenMP library to divide the processing of the prenex formula into independent threads of execution, each running on a separate Intel Xeon CPU. Each of these threads runs an instance of QEPCAD B. The final results are averaged over 50 different randomly generated prenex formulae.\\\\

The comparison between the running times of QEPCAD B with the preprocessor and without are shown in Table 1 and Fig. 6 compares the running times (with and without the preprocessor) of QEPCAD B over 50 different prenex formulae in two variables.

\begin{table*}
\begin{tabular}{| c | c | c | c | c |}
\hline
  Variables & No. of terms per polynomial & No. of polynomials & Time without the preprocessor & Time with the preprocessor \\
\hline
\hline
  2 & 5 & 5 & 3 Sec & 0.3 Sec \\
\hline
  3 & 5 & 5 & $\infty$ & 92 Sec \\
\hline
  4 & 2 & 2 & 77000 Sec & 2 Sec \\
\hline
\end{tabular}
\begin{center}
\caption{Comparison of running times of QEPCAD B with the proposed preprocessor and without.}
\end{center}
\end{table*}

\begin{figure*}
\begin{center}
  \includegraphics[scale=0.5]{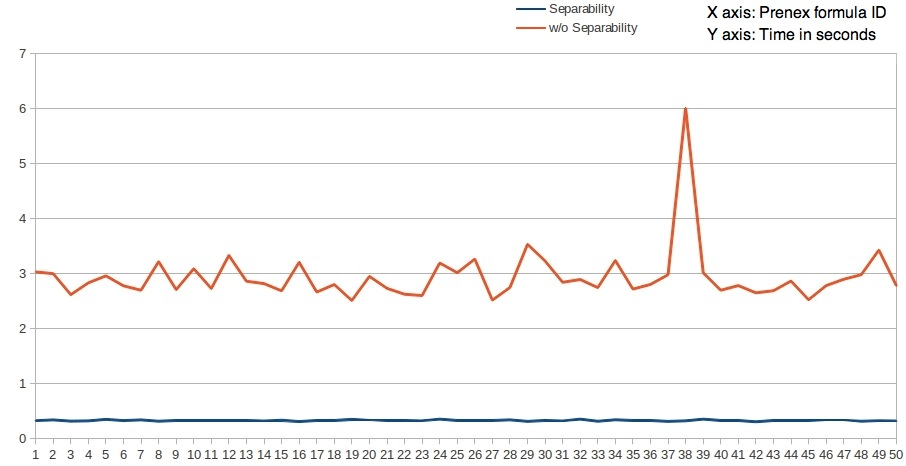}\\
\caption{The comparison between the QEPCAD B with preprocessor and without the preprocessor: A 2 variable case study over 50 randomly generated prenex formulae}
  \includegraphics[scale=0.5]{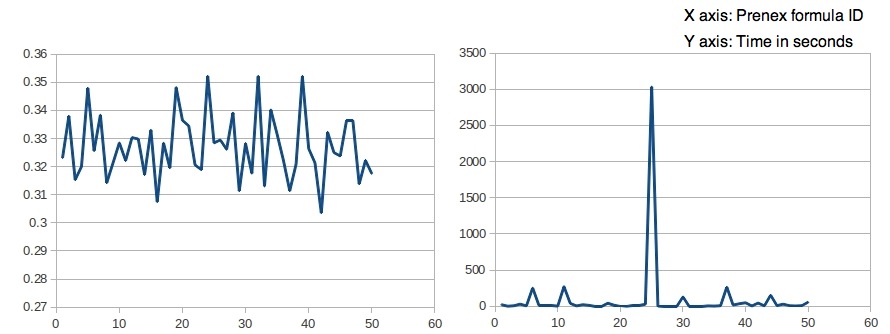}\\
\caption{The behaviour of QEPCAD B with the preprocessor for prenex formulae with 2 variables (left) and prenex formulae with 3 variables (right).}
\end{center}
\end{figure*}
An example of such a randomly generated formula in 2 variables from the separable class $S_{10}$, each with at most 5 polynomials and each polynomial with at most 5 terms is,

$(\forall x0)[[[(58) {x_{0}}^{8} {x_{1}}^{2}  + (-68) {x_{0}}^{10} {x_{1}}^{4}  + (73)  + (-4)  + (-28) {x_{0}}^{9}  = 0] \vee[(44) {x_{0}}^{4} {x_{1}}^{5}  + (-88) {x_{0}}^{10} {x_{1}}^{4}  + (11) {x_{1}}^{9}  + (-70)  + (-19) {x_{0}}^{6} {x_{1}}^{5}  = 0] \vee[(20) {x_{0}}^{7}  + (-61) {x_{0}}^{10} {x_{1}}^{3}  + (-71)  + (81) {x_{0}}^{3} {x_{1}}^{5}  + (14) {x_{0}}^{7}  = 0] \vee[(-58) {x_{1}}^{5}  + (-35) {x_{0}}^{10}  + (5)  + (30)  + (45) {x_{1}}^{5}  = 0] \vee[(-69) {x_{0}}^{5} {x_{1}}^{6}  + (71) {x_{0}}^{10}  + (40)  + (5)  + (-90) {x_{1}}^{5}  = 0]] \wedge [[(-67) {x_{0}}^{2} {x_{1}}^{6}  + (-17) {x_{0}}^{10}  + (-18) {x_{1}}^{3}  + (14)  + (37) {x_{0}}^{6}  = 0] \vee[(76) {x_{0}}^{5} {x_{1}}^{1}  + (-7) {x_{0}}^{10}  + (-44) {x_{0}}^{5} {x_{1}}^{6}  + (-96) {x_{0}}^{5}  + (-69) {x_{0}}^{1} {x_{1}}^{6}  = 0] \vee[(-73)  + (-42) {x_{0}}^{10}  + (67) {x_{0}}^{7} {x_{1}}^{2}  + (-50) {x_{1}}^{9}  + (-12) {x_{1}}^{1}  = 0] \vee[(63) {x_{0}}^{2} {x_{1}}^{7}  + (23) {x_{0}}^{10}  + (-74) {x_{1}}^{5}  + (31)  + (17) {x_{0}}^{7}  = 0] \vee[(-47) {x_{0}}^{1} {x_{1}}^{9}  + (-31) {x_{0}}^{10} {x_{1}}^{1}  + (17)  + (17)  + (-88) {x_{1}}^{6}  = 0]] \wedge [[(78) {x_{0}}^{9}  + (-43) {x_{0}}^{10}  + (5) {x_{0}}^{6}  + (4) {x_{1}}^{7}  + (-76)  = 0] \vee[(-23) {x_{0}}^{3}  + (81) {x_{0}}^{10}  + (36)  + (35)  + (54) {x_{1}}^{1}  = 0] \vee[(-29) {x_{0}}^{5}  + (54) {x_{0}}^{10} {x_{1}}^{2}  + (-20) {x_{0}}^{6} {x_{1}}^{1}  + (57) {x_{0}}^{2}  + (62) {x_{0}}^{9} {x_{1}}^{4}  = 0] \vee[(77) {x_{1}}^{6}  + (97) {x_{0}}^{10} {x_{1}}^{6}  + (-42)  + (-11) {x_{0}}^{1}  + (93) {x_{0}}^{3} {x_{1}}^{4}  = 0] \vee[(37) {x_{0}}^{7}  + (1) {x_{0}}^{10}  + (-18) {x_{0}}^{1} {x_{1}}^{9}  + (36)  + (25) {x_{0}}^{6}  = 0]] \wedge [[(-35) {x_{0}}^{3}  + (22) {x_{0}}^{10}  + (-80) {x_{0}}^{3}  + (13) {x_{1}}^{3}  + (96) {x_{1}}^{5}  = 0] \vee[(-73)  + (-25) {x_{0}}^{10} {x_{1}}^{5}  + (79)  + (-16)  + (88) {x_{0}}^{3} {x_{1}}^{7}  = 0] \vee[(-94) {x_{0}}^{8} {x_{1}}^{6}  + (86) {x_{0}}^{10}  + (-52) {x_{0}}^{2}  + (15)  + (32) {x_{0}}^{2}  = 0] \vee[(49)  + (-36) {x_{0}}^{10}  + (51) {x_{0}}^{5} {x_{1}}^{8}  + (59) {x_{0}}^{4}  + (-37)  = 0] \vee[(99) {x_{0}}^{8}  + (71) {x_{0}}^{10}  + (73) {x_{1}}^{5}  + (68) {x_{0}}^{8} {x_{1}}^{3}  + (51) {x_{0}}^{1} {x_{1}}^{1}  = 0]] \wedge [[(70)  + (-53) {x_{0}}^{10} {x_{1}}^{5}  + (73) {x_{0}}^{7} {x_{1}}^{9}  + (61) {x_{1}}^{3}  + (-59) {x_{0}}^{3} {x_{1}}^{9}  = 0] \vee[(-32) {x_{0}}^{3} {x_{1}}^{5}  + (-80) {x_{0}}^{10}  + (-28)  + (-88) {x_{0}}^{4}  + (35) {x_{0}}^{1}  = 0] \vee[(-65)  + (-81) {x_{0}}^{10}  + (35) {x_{0}}^{9} {x_{1}}^{6}  + (8) {x_{0}}^{7} {x_{1}}^{4}  + (-38) {x_{0}}^{3}  = 0] \vee[(-24) {x_{1}}^{6}  + (26) {x_{0}}^{10}  + (15) {x_{0}}^{1}  + (80) {x_{1}}^{6}  + (93)  = 0] \vee[(-42) {x_{0}}^{1}  + (84) {x_{0}}^{10}  + (-13) {x_{0}}^{5}  + (33)  + (-17) {x_{0}}^{5} {x_{1}}^{7}  = 0]] \wedge [[(75) {x_{0}}^{8}  + (26) {x_{0}}^{10} {x_{1}}^{7}  + (-47) {x_{0}}^{5}  + (8)  + (-81) {x_{0}}^{8} {x_{1}}^{9}  = 0] \vee[(-96) {x_{0}}^{4}  + (24) {x_{0}}^{10} {x_{1}}^{8}  + (-78)  + (82) {x_{1}}^{6}  + (-22) {x_{0}}^{4}  = 0] \vee[(69) {x_{1}}^{5}  + (12) {x_{0}}^{10} {x_{1}}^{3}  + (-9) {x_{1}}^{2}  + (63) {x_{0}}^{1}  + (-39) {x_{0}}^{2}  = 0] \vee[(-27)  + (1) {x_{0}}^{10}  + (44) {x_{0}}^{9} {x_{1}}^{4}  + (-68) {x_{0}}^{8} {x_{1}}^{2}  + (-3)  = 0] \vee[(94) {x_{1}}^{3}  + (-85) {x_{0}}^{10} {x_{1}}^{2}  + (-63)  + (22) {x_{1}}^{6}  + (-74) {x_{1}}^{8}  = 0]] \wedge [[(-97)  + (-7) {x_{0}}^{10}  + (27) {x_{0}}^{8} {x_{1}}^{7}  + (71)  + (-26) {x_{0}}^{9} {x_{1}}^{1}  = 0] \vee[(15) {x_{1}}^{8}  + (13) {x_{0}}^{10}  + (-15)  + (9)  + (16) {x_{0}}^{3} {x_{1}}^{2}  = 0] \vee[(7) {x_{1}}^{6}  + (13) {x_{0}}^{10} {x_{1}}^{1}  + (7) {x_{0}}^{9}  + (-87)  + (13) {x_{1}}^{3}  = 0] \vee[(-8)  + (-27) {x_{0}}^{10} {x_{1}}^{6}  + (-44) {x_{0}}^{3} {x_{1}}^{4}  + (-66) {x_{0}}^{2} {x_{1}}^{9}  + (42) {x_{1}}^{3}  = 0] \vee[(-13)  + (-95) {x_{0}}^{10} {x_{1}}^{6}  + (48) {x_{0}}^{3} {x_{1}}^{7}  + (8) {x_{1}}^{3}  + (-95) {x_{0}}^{5}  = 0]] \wedge [[(-51) {x_{0}}^{5}  + (-40) {x_{0}}^{10} {x_{1}}^{3}  + (-18) {x_{0}}^{6}  + (22) {x_{1}}^{2}  + (-65)  = 0] \vee[(-22) {x_{0}}^{7}  + (-73) {x_{0}}^{10} {x_{1}}^{5}  + (19) {x_{0}}^{9}  + (84) {x_{0}}^{7}  + (43) {x_{0}}^{8}  = 0] \vee[(39) {x_{1}}^{8}  + (95) {x_{0}}^{10} {x_{1}}^{3}  + (-57) {x_{1}}^{2}  + (-1) {x_{1}}^{3}  + (21)  = 0] \vee[(21) {x_{1}}^{3}  + (-80) {x_{0}}^{10}  + (-14)  + (56) {x_{1}}^{5}  + (4) {x_{0}}^{4} {x_{1}}^{9}  = 0] \vee[(-56)  + (-88) {x_{0}}^{10} {x_{1}}^{1}  + (90) {x_{1}}^{9}  + (31)  + (-63) {x_{0}}^{2}  = 0]] \wedge [[(73) {x_{1}}^{3}  + (91) {x_{0}}^{10}  + (79) {x_{0}}^{9}  + (72) {x_{0}}^{8}  + (-97) {x_{1}}^{4}  = 0] \vee[(-36) {x_{0}}^{8}  + (2) {x_{0}}^{10} {x_{1}}^{5}  + (-46) {x_{0}}^{7} {x_{1}}^{2}  + (11)  + (28) {x_{0}}^{7} {x_{1}}^{7}  = 0] \vee[(-38) {x_{0}}^{5}  + (94) {x_{0}}^{10} {x_{1}}^{9}  + (-15) {x_{0}}^{1}  + (-91)  + (-5) {x_{1}}^{3}  = 0] \vee[(93)  + (52) {x_{0}}^{10} {x_{1}}^{2}  + (-5) {x_{0}}^{3} {x_{1}}^{2}  + (-20)  + (-5)  = 0] \vee[(-47) {x_{0}}^{7}  + (80) {x_{0}}^{10}  + (76)  + (54) {x_{1}}^{3}  + (86) {x_{1}}^{4}  = 0]] \wedge [[(-21) {x_{0}}^{6} {x_{1}}^{8}  + (-83) {x_{0}}^{10} {x_{1}}^{5}  + (67) {x_{0}}^{3}  + (80) {x_{0}}^{5}  + (57) {x_{1}}^{2}  = 0] \vee[(-24) {x_{1}}^{6}  + (78) {x_{0}}^{10}  + (-68) {x_{0}}^{6}  + (83) {x_{0}}^{8}  + (66)  = 0] \vee[(-60)  + (97) {x_{0}}^{10} {x_{1}}^{7}  + (-6) {x_{0}}^{3}  + (53) {x_{1}}^{9}  + (-36) {x_{1}}^{4}  = 0] \vee[(83) {x_{0}}^{1} {x_{1}}^{5}  + (-55) {x_{0}}^{10} {x_{1}}^{6}  + (-48)  + (69) {x_{0}}^{2}  + (-44)  = 0] \vee[(15)  + (-49) {x_{0}}^{10}  + (90)  + (-23) {x_{0}}^{6} {x_{1}}^{4}  + (85) {x_{1}}^{4}  = 0]]].$

An example of the actual input, which consists of 100 clauses could not be provided here due to space constraints (as it would occupy four pages). This also proves the robustness of the algorithm, as CAD in practice relies on the systematic factorization of the polynomials. Using random formulae demonstrates the applicability of the algorithm even if the probability of such factorization is 0.

The behaviour of the running time of QEPCAD B with the preprocessor for 50 randomly generated prenex formulae is depicted in Fig. 7.


\subsection{On Effectiveness of the Sharing Factor}

This experiment aims to study the sharing factor and the impact it can have on the amount of computational work done by the CAD algorithm. The space utilized for execution is taken as a metric for computational work. The number of cells utilized is provided by the QEPCAD B tool. A formula from the separable class $S_2$ is considered, with 2 polynomials per clause, 2 terms per polynomial and 6 variables. The sharing factor is varied from 0 to 3. The first clause is kept constant for all the cases and the second clause is varied. The second clause uses $k$ variables, out of those used in the first clause, for a sharing factor of $k$.

QEPCAD B is run twice for each sharing factor, the first run being the whole formula and the first clause being truncated in the second run. Any structure imposed by the first clause on the second clause should appear as the difference between the space utilized by both the runs. As the same clause is deleted in all 4 cases, the experiment should not react to factors other than the sharing factor. As shown in Table 2, in the 0 sharing factor case (which is from $C_2$), the space utilized in both cases is almost identical. The difference increases as the sharing factor increases. The case with sharing factor of 1 is taken as an anomaly, where there is a greater amount of dependence on the one variable that is shared. This demonstrates the existence of corner cases.

As an example, we provide the formula used for the experiment concerning sharing factor of 3.

$(\forall x_0)(\forall x_1)(\forall x_2)(\forall x_3)[[[(85) x_1^1 x_2^2  + (64) x_0^3 x_2^1  = 0] \vee [(41) x_0^1 x_2^1  + (-96) x_0^3 x_1^2 x_2^2  = 0]] \wedge [[(-18) x_4^1  + (-44) x_0^3 x_2^1 x_4^2 x_5^1  = 0] \vee [(-78) x_1^1 x_3^1 x_4^1 x_5^2  + (31) x_0^3 x_1^1 x_2^2 x_4^2 x_5^2  = 0]]]$

The variables $x_0,x_1,x_2$ are shared among the two clauses in this example. The first clause is common to all the four cases studied.

\begin{table}
\begin{tabular}{| p{2.5cm} | p{2cm} | p{2cm} |}
\hline
  \textbf{Sharing Factor} & \textbf{Percentage of space utilized (with the first clause)} & \textbf{ Percentage of Space utilized (without the first clause)} \\
\hline
\hline
  0 shared variables & 7.6\%(38187 cells) & 3.7\%(18723 cells)\\
\hline
  1 shared variables & 44.18\%(279058 cells) left after 49 garbage collections & 5.14\%(25735 cells) \\
\hline
  2 shared variables & 69.31\%(346561 cells) & 3.32\%(16614 cells)\\
\hline
  3 shared variables & 65.4\%(327124 cells) left after 2 garbage collections & 5.04\%(25234 cells) \\
\hline

\end{tabular}
\begin{center}
\caption{An analysis of the effectiveness of sharing factor to study the structure of a prenex formula}
\end{center}
\end{table}

\section{Concluding Remarks}

In this paper, a preprocessor to CAD is proposed based on the notions of clause normal forms and virtual substitutions. This preprocessor uses clause normal forms to impose a structure on an input formula and runs several instances of CAD on these clauses. The effectiveness of this preprocessor is studied experimentally and some theoretic notions of structure inherent in a formula and the consequences of losing it are presented. The paper ends by providing the notion of an `idealistic' scenario and a measure of deviation that a random prenex formula has from it.

Though the notion showing factor reflects parallelizability of CAD to some extent it cannot completely describe the best-case scenario for the preprocessor.  We seek to improve upon this notion by considering parameters other than just the sharing factor. This will help us to theoretically analyze the time and space complexity of CAD with the proposed preprocessor.

\bibliographystyle{plain}
\bibliography{harikrishnamalladi}

\end{document}